\documentstyle[aps,psfig,epsfig]{revtex}

\newcommand{\ave}[1]{\mbox{$\langle #1 \rangle$}}

\begin{document}

\title{Scaling behavior of randomly alternating surface growth processes}
\author{Subhadip Raychaudhuri$^{(1)}$, Yonathan
Shapir$^{(1,2)}$}\address{$^{(1)}$Department of Physics and
Astronomy, University of Rochester, Rochester, NY 14627, USA}
\address{$^{(2)}$Department of Chemical Engineering,
University of Rochester, Rochester, NY 14627, USA}

\date{\today}
\maketitle

\begin{abstract}

The scaling properties of the roughness of surfaces grown by
two different processes randomly alternating in time, are addressed.
The duration of each application of the two primary processes is
assumed to be independently drawn from given distribution functions.
We analytically address processes in which the two primary
processes are linear and extend the conclusions
to non-linear processes as well. The growth scaling
exponent of the average roughness with the number of applications
is found to be determined by the long time tail of the
distribution functions. For processes in which both mean application
times are finite, the scaling behavior follows that of the
corresponding cyclical process in which the uniform application
time of each primary process is given by its mean.
If the distribution functions decay with a small enough  power law
for the mean application times to diverge, the growth
exponent is found to depend continuously on this power law exponent.
In contrast, the roughness exponent does not depend on the
timing of the applications. The analytical results are supported
by numerical simulations of various pairs of primary processes and
with different distribution functions. Self-affine surfaces
grown by two randomly alternating processes are common in nature
(e.g., due to randomly changing weather conditions) and in
man-made devices such as rechargeable batteries.
\end{abstract}

\section{INTRODUCTION}
\subsection{Motivation and background}
The scaling approach to kinetic roughening of surfaces has been
found to be very useful in a large number of systems  [1-7]. It is
based on their self-affine property, reflected by the power-law
dependence of the roughness on the system size in the, steady
state, large-time regime.
In the early regime, the roughness also increases as a power law
of the time. Systems in which such a behavior of the surface
roughness was observed and analyzed include different (vapor,
electrochemical, epitaxial, etc.) deposition processes, burning
front, malignant tumors, and others. Similarly, self-affine rough
surfaces are formed in opposite processes of recession,
desorption, dissolution, etc.

Very recently we have addressed the scaling behavior of cyclical
processes in which two different processes (typically one growth
and one recession) follow each other regularly. Such cyclical
processes are abundant in natural and artificial systems. Many
natural growth processes are affected by changing degree of
illuminations through the daily cycle or seasonal variations in
weather conditions. Rechargeable batteries, charged and discharged
periodically, provide an example of a practical application in
which a rough metal surface accumulates on one of the electrode
(and cause a short when it reaches the other electrode).
Chemotherapeutic (or radiation) treatment of a malignant tumor
applied periodically and the expansion/curtailment of a bacterial
colony depending on the availability of the nutrient,  may provide
examples in the biosciences. Experiments on the roughness of
silver surface grown by cyclical electrodeposition/dissolution
were found to be consistent with the scaling predictions. The main
results of our theoretical investigations are summarized in Sec.
II.

While studying these cyclical processes it was also realized that
many of them are not exactly periodic. Namely the duration of one
(or both) primary process, may not be uniform. For example, the
charge and discharge time of a battery maybe very irregular
depending on the way it is utilized. If a growth process depends
on atmospheric conditions such as alternating rain and sunshine,
the length of the application time of each primary process will
also fluctuate drastically. That may be the case for alternating
dry/wet erosion processes of a natural rock bed or a man-built
wall.

So two crucial questions which ought to be addressed are: To what
extent the cyclical theory applies to systems with non-uniform
application times of the primary processes? And if the cyclical
theory fails to describe them, how are their scaling properties
modified? The goal of the present work is to seek and provide the
answers to these two essential questions.

\subsection{Basic scaling concepts and relations}

The roughness of self-affine surfaces created by growth or
recession processes have been described using scaling concepts
\cite{family2}. The width $W(L,t)$ of a surface linear dimension
$L$ at a time $t$ is given by

\begin{equation} \label{eq:}
 W(L,t) = \Bigl \langle \, \Bigl( h( \vec{r},t) - \langle h(\vec{r},t)
   \rangle \Bigr)^{2} \,  \Bigr \rangle   ^{\frac{1}{2}},
\end{equation}

where the angular bra-ket $\langle \rangle$ denotes average over
both lateral sites and different realizations of surface
configurations. $W(L,t)$ was found to scale as \cite{family2}
\begin{equation} \label{eq:}
 W(L,t) \sim  L^{\alpha} \,  g(L/ \xi(t)),
\end{equation}
where $\xi(t) \sim t^{1/z}$ is the lateral correlation length. $g$
is the scaling function which behaves such that for large time $(t
\gg L^{z})$  $W \sim L^{\alpha}$ ($\alpha$ is the roughness
exponent), while in the early-time regime
($t \ll L^{z}$),  $W \sim t^{\beta}$
(where $\beta = \alpha /z$ is the growth exponent). Different
surfaces are classified into universality classes which share the
same set of scaling exponents.  Some generic universality classes
are mentioned in subsection C.

For rough surfaces formed by cyclical processes a suitable scaling
law in terms of the number of cycles $n$ was introduced
\cite{shapir}\cite{raychaudhuri}. The width $W(L,n)$ for a system
of linear size $L$ obeys the following scaling form:

\begin{equation} \label{eq:}
             W(L,t) = L^{\alpha} \, g_{c}(L/\xi_{c}(n)),
\end{equation}

where $\xi_{c}(n) \sim n^{1/z}$ is the lateral correlation length
and $g_{c}$ is the cyclical scaling function.
\newline For large time \hspace{0.1in}
$n \gg L^{z}$: \hspace{0.4in} $W \sim L^{\alpha}$,
\newline while for \hspace{0.5in}
$n \ll L^{z}$:  \hspace{0.4in} $W \sim n^{\beta}$,
\newline where $\beta = \alpha /z$ is the growth exponent.
Again, different universality classes can be defined depending on
the values of the exponents $\alpha$ and $\beta$. In most of the
cases studied by us, the scaling exponents of the cyclical process
were identical to those of one of the primary processes (the
so-called dominating process).

In our previous papers on cyclical growth \cite{shapir}
\cite{raychaudhuri} the durations of the primary processes (in
every cycle) were assumed to be uniform. However, in many
realistic situations the durations of the primary processes will
vary from cycle to cycle.

\subsection{Growth Models and Universality Classes}

The growth processes fall into different universality
classes\cite{bara}. All processes within one class share the same
exponents and their asymptotic continuum stochastic equations
differ at most by irrelevant terms (in the renormalization group
(RG) sense). Using the symbolic  index $i=1,2,..$ to denote
different  processes, the ones we consider here follow growth
equations of the form:

\begin{equation} \label{eq:2}
\frac{\partial h(\vec{r},t)}{\partial t} = A_{i} \{h\} +
\eta_{i}(\vec{r},t) + v_{i},
\end{equation}
where $A_{i}$\{$h\}$ is a local functional depending on the
spatial derivatives of $h(\vec{r},t)$ and the noise
$\eta_{i}(\vec{r},t)$ reflects the the random fluctuations in the
deposition process and satisfies

\begin{equation} \label{eq:}
       \langle \eta_{i}(\vec{r},t) \rangle = 0,
\end{equation}

and

\begin{equation} \label{eq:}
\langle \eta_{i}(\vec{r}_{1},t_{1}) \, \eta_{i}(\vec{r}_{2},t_{2})
\rangle = 2D_{i} \delta ^{\tilde d} (\vec{r}_{1}-\vec{r}_{2}) \,
\delta(t_{1}-t_{2}),
\end{equation}
where $\tilde d = d-1$ is the substrate dimension.

Some of the generic growth processes we will consider are:

(i) Linear: Random deposition (RD), Edwards-Wilkinson (EW)
\cite{edwards} and Mullins-Herring (MH) \cite{MH}\cite{dasarma}
\cite{wolf}.

(ii) Nonlinear: Kardar-Parisi-Zhang (KPZ) \cite{kardar} and
Molecular Beam Epitaxy (MBE) \cite{laid} \cite{vill}.

\subsection{The structure factor (SF)}

The structure factor is defined as $S(\vec{q},t) = \bigl \langle
h(\vec{q},t) \, h(\vec{-q},t) \bigr \rangle$, where $h(\vec{q},t)$
is the Fourier transform of the height $h(\vec{r},t)$. In the
theoretical analysis of surface growth it is convenient to work in
Fourier space and compute first the SF rather than the roughness
$W$ itself. Also, in some experiments the surface is probed by
scattering processes which provide its structure factor.

Dynamic Scaling hypothesis Eq. (2) can be translated to the
Fourier space such that \cite{bara}
\begin{equation} \label{eq:}
   S(\vec{q},t) = q^{-\tilde d -2\alpha} \, g(q/t^{-1/z}).
\end{equation}
Surface width $W$ can be readily calculated from $S(\vec{q},t)$
using the relation $W^{2}(L,t) = (1/L^{\tilde d}) \sum_{\vec{q}}
S(\vec{q},t)$.

The organization of the rest of the paper is as follows: In the
next section (II) we briefly review the theoretical analysis of
cyclical processes and their main results.In Section III Some
exact results are derived for the case of two linear primary
processes with non-uniform application times. In Section IV
various probability distributions of the random application times
of the primary processes are considered. Results of numerical
simulations for both linear and nonlinear primary processes are
presented. Sec. VI contains a summary of our results.

\section{Review of Cyclical Growth Processes}

Any cyclical growth process consist of two primary processes.  The
durations for the first and the second processes are $T_{1}=pT$
and $T_{2}=(1-p)T$, respectively. The period of one cycle is $T =
T_{1} + T_{2}$. The cyclic growth equation (equivalent to Eq. (4)
of a simple growth/recession) in terms of the basic two processes
is

\begin{equation}\label{eq:}
\frac{\partial h}{\partial t}=[a_{1}h + \eta _{1}
 + v_{2}]\Theta(p-f(t)) +
[a_{2}h + \eta_{2} + v_{2}]\Theta(f(t)-p),
\end{equation}

where $f(t)$ is defined as the fractional part of $t/T$ and
$\Theta(x)$ is the unit step function.

\subsection{Linear Primary Processes}

When both the primary processes are linear and their durations
($T_{1}$ and $T_{1}$) are constants the above Eq. (8) can be
solved exactly
to yield the SF. These results are is briefly described below and
details may be found in our previous papers \cite{shapir}
\cite{raychaudhuri}.

For a linear process Langevin equation of the form of Eq. (4)  can
be easily solved in Fourier space to yield the SF

\begin{equation} \label{eq:}
S(q,t) = \exp$\{$-2a(q)t$\}$ S(q,0) +
         \frac{D}{a(q)}[1-\exp(-2a(q)t)],
\end{equation}

where $S(q,0)$ is the SF at $t=0$ which contains the information
of the initial roughness. We assume $S(q,0)$ to be zero, {\it
i.e.} the growth starts from a flat substrate. This will not
affect the asymptotic scaling exponents.

We index the two primary processes by $i=1,2$ respectively and
define $\bar{a_{i}} = a_{i}(q)T_{i}$. During the $1st$ cycle of
the cyclical growth, the structure factor generated by the first
primary process (of duration $T_{1}= pT$) is assigned as the
initial condition for the second primary process. The second
process lasts for $T_{2}=(1-p)T$ to yield the structure factor
$S_{c}(q,1)$ of the cyclical process after the first cycle. This
is again used as the initial structure factor for the first
process in the $2nd$ cycle. If we keep iterating like this, the SF
after $n$ cycles becomes

\begin{eqnarray} \label{eq:}
S_{c}(q,n) & = & \bigg{[} \frac{D_{1}}{a_{1}}\exp(-2a_{2}T_{2})
\left(1-\exp(-2a_{1}T_{1}) \right)
\nonumber \\
& & \quad + \frac{D_{2}}{a_{2}}\left(1-\exp(-2a_{2}T_{2})\right)
\bigg{]} \left[{1-\exp(-2a_{c}n) \over1-\exp(-2a_{c})} \right].
\end{eqnarray}

where $\bar a_{c} = a_{c}T$ with

\begin{equation} \label{eq:}
         a_{c} = [a_{1}p + a_{2}(1-p)].
\end{equation}

 In the scaling limit of small $q$, Eq. () for the cyclical SF reduces to

\begin{equation} \label{eq:}
S_{c}(q,n) \sim \frac{D_{c}}{a_{c}(q)} \left[1-\exp
\big(-2a_{c}(q) \, T \, n \big) \right],
\end{equation}

where the effective noise strength for the cyclic process is
defined as

\begin{equation} \label{eq:}
      D_{c}=pD_{1}+(1-p)D_{2}.
\end{equation}

In terms of the effective cyclic parameters $a_{c}$ and $D_{c}$,
the SF (12) for the cyclical process looks very similar to that of
a generic linear growth process (see Eq. (9)) if the time variable
is replaced by the {\it number of cycles} $n$. Hence, the dynamic
scaling analysis (Eq. (7)) of the structure factor can be used to
determine the scaling exponents in the case of a cyclical growth.

In the regime of saturated roughness, after large number of cycles
$n \rightarrow \infty$, Eq. (12) yields $S_{c}(q,n) \sim
\frac{D_{c}}{a_{c}(q)}$. The roughness exponent of the cyclic
process is determined by the $q \rightarrow 0$ divergence of
$\frac{1}{a_{c}(q)}$. Since $a_{c} = [a_{1}p + a_{2}(1-p)]T$ and
$a_{i}(q) \sim q^{z_{i}}$, it is the process with smaller $z_{i}$
which dominates the asymptotic cyclical roughness. In the growing
phase of the interface roughness, the number of cycles $n$ is
multiplied by $a_{c}(q)$ in Eq. (12). The process with smaller
$z_{i}$ will again dominate in the $q \rightarrow 0$ asymptotic
limit. Therefore, the primary process with the smaller dynamic
exponent carries over its scaling exponents to the combined
cyclical  process.

\subsection{Non-linear primary processes}

We extended our analysis to non-linear processes as well by using
an approximate RG approach. In this approach we ``set aside'' the
non-linear terms for the first few RG iterations in which all
fluctuations on time scales smaller than the time one full cycle
are integrated out. These few initial steps deals only with the
linear parts of the two primary processes and result in an
effective linear process, as described above. At this stage the
non-linear terms are added back to the stochastic equation with
their bare couplings multiplied by $p$ (respectively, $1-p$) to
take into account the fraction of time of their operation. The
next iterations of the RG process then may be applied in the usual
manner. This approach assumes that the approximation made in the
few initial iterations will not change the ultimate fixed-points
to which the RG flow take the system. We expect this to be the
case if the system location in parameter space is not too close to
a separatrix between two basins of attraction.

\section{Analytical considerations of randomly alternating processes}

\subsection{Linear primary processes}

Now we are ready to consider the more general case of nonuniform
durations of the two primary processes. The time elapsed during
the first and the second processes in the $k^{th}$ application are
denoted by $t_{1}^{(k)}$ and  $t_{2}^{(k)}$ respectively. So
$t_{1}^{(k)}$ and  $t_{2}^{(k)}$ are assumed to be random
variables obeying specific probability distribution functions
({\it pdf}). We also define two new random variables $\xi_{k} =
exp[ - 2 a_{1}(q) t_{1}^{(k)}]$ and $\eta_{k} = exp[ - 2 a_{2}(q)
t_{2}^{(k)}]$, in terms of which the structure factor of the
cyclic process will be expressed.

The SF, after the first primary process has taken place in the
first cycle, turns out to be

\begin{equation} \label{eq:}
    S(q,T_{1}^{(1)}) = \frac{D_{1}}{a_{1}(q)}[1-\exp(-2a_{1}(q)t)]
                     = \frac{D_{1}}{a_{1}(q)} (1 - \xi_{1}),
\end{equation}

where $S(q,0)=0$ is assumed for simplicity. Hence, after one
complete cycle the SF becomes

\begin{equation} \label{eq:}
    S_{c}(q,1) =  (\frac{D_{1}}{a_{1}})\eta_{1}(1 - \xi_{1}) +
             (\frac{D_{2}}{a_{2}})(1 - \eta_{1}),
\end{equation}

which can now be used as the initial SF for the first primary
process of the next cycle. After the end of the $2^{nd}$ cycle,
the SF is

\begin{equation} \label{eq:}
 S_{c}(q,2) = (\frac{D_{1}}{a_{1}}) \left \{ \xi_{2}\eta_{1}\eta_{2}
(1-\xi_{1}) + \eta_{2}(1-\xi_{2}) \right \} +
(\frac{D_{2}}{a_{2}}) \left \{ \xi_{2}\eta_{2}(1-\eta_{1}) +
(1-\eta_{2}) \right \}.
\end{equation}

Proceeding in this manner we finally arrive at the general
expression of the SF after $n$ cycles

\begin{eqnarray} \label{eq:}
 S_{c}(q,n) = (\frac{D_{1}}{a_{1}}) \left \{
\sum_{r=1}^{n-1} (\prod_{i=r+1}^{n}\xi_{i+1}
\prod_{j=r}^{n}\eta_{j}) (1 - \xi_{r}) +
                \eta_{n} (1 - \xi_{n}) \right \}
\nonumber \\
+(\frac{D_{2}}{a_{2}}) \left \{ \sum_{r=1}^{n-1}
(\prod_{i=r+1}^{n} \xi_{i} \prod_{j=r+1}^{n}\eta_{j}) (1 -
\eta_{r}) +
                         (1 - \eta_{n}) \right \}
\end{eqnarray}
(Note the similarity of this sum of products of random variables
to that encountered in 1D diffusion in random environment
\cite{bouchaud}. For that system a similar expression yields the
first passage time and the random terms in the products are the
ratios of the transition probabilities in the two directions
between neighboring sites). Suppose the duration of the primary
processes $t_{1}^{(k)}$ and $t_{2}^{(k)}$ are distributed with the
{\it pdf} $P(t_{1})$ and $P(t_{2})$, respectively. We need to take
the average with respect to these $P(t_{i})$ of the SF in order to
obtain the scaling behavior of the roughness. However, the above
expression of the SF contains various products of the random
variables $\xi$ and $\eta$ and in such situations the average
behavior might differ from the typical behavior. Since by
definition $\xi_{k} = exp[ - 2 a_{1}(q) t_{1}^{k}]$ and $\eta_{k}
= exp[ - 2 a_{2}(q) t_{2}^{k}]$, they always take values less than
one. Thus, $\ave{\xi^{m}} < 1$ and $\ave{\eta^{m}} < 1$
($m=1,2....$), {\it i.e.} all the moments including the averages
remain less than unity, and also $\ln \ave{e^{-2a_{i}t_{i}^{k}}} <
0$. This is valid for any normalizable probability distribution
due to the fact that the durations are always positive ($t_{i}^{k}
\ge 0$). As a result, the SF is not dominated by the higher
products of $\xi_{k}$ and $\eta_{k}$, rather the most dominant
contribution in the sum will come from the terms linear in the
random variables. In that case, the fluctuation around the average
will be small and the typical behavior will be well represented by
the average. By taking the average of the Eq. (17)

\begin{eqnarray} \label{eq:}
\ave{S_{c}(q,n)} & = & (\frac{D_{1}}{a_{1}}) \bigg{<}
\sum_{r=1}^{n-1} \left(\prod_{i=r+1}^{n}\xi_{i+1}
\prod_{j=r}^{n}\eta_{j}\right) (1 - \xi_{r}) +
                \eta_{n} (1 - \xi_{n})  \bigg{>}
\nonumber \\
& & \quad + (\frac{D_{2}}{a_{2}}) \bigg{<} \sum_{r=1}^{n-1}
\left(\prod_{i=r+1}^{n} \xi_{i} \prod_{j=r+1}^{n}\eta_{j} \right)
(1 - \eta_{r}) +
                         (1 - \eta_{n})  \bigg{>}
\nonumber \\
& = &  (\frac{D_{1}}{a_{1}}) \left \{ \sum_{r=1}^{n-1}
(\prod_{i=r+1}^{n} \ave{\xi} \prod_{j=r}^{n}\ave{\eta}) (1 -
\ave{\xi}) +
                \ave{\eta} (1 - \ave{\xi}) \right \}
\nonumber \\
& & \quad + (\frac{D_{2}}{a_{2}}) \left \{ \sum_{r=1}^{n-1}
(\prod_{i=r+1}^{n} \ave{\xi} \prod_{j=r+1}^{n}\ave{\eta}) (1 -
\ave{\eta}) +
                         (1 - \ave{\eta}) \right \},
\end{eqnarray}

where $\ave{\xi} = \ave{\exp[-2a_{1}t_{1}^{i}]}$ and $\ave{\eta} =
\ave{\exp[-2a_{2}t_{2}^{i}]}$. The above expression (18) of the SF
has the form of a geometric series which can be easily summed with
the result

\begin{equation} \label{eq:}
\ave{S_{c}(q,n)} = \bigg {\{} (\frac{D_{1}}{a_{1}}) \ave{\eta}
(1-\ave{\xi}) + (\frac{D_{2}}{a_{2}})  (1-\ave{\eta}) \bigg {\}}
\bigg {[} \frac{1-\ave{\xi}^{n}\ave{\eta}^{n}}
{1-\ave{\xi}\ave{\eta}} \bigg {]}
\end{equation}

This form of the average SF resembles the SF ( Eq. (10)) for the
case of uniform durations of the primary processes. In fact, one
can easily arrive at Eq. (10) starting from Eq.(19), once the
uniformity of $t_{1}^{k}$ and $t_{2}^{k}$ are assumed, i.e. $P(t)
= \delta (t-T)$.

The scaling exponents in this case of random application times may
be extracted using the small $q$ divergence (as done in the
uniform case also) of the average SF (19), given in terms of
$\ave{\xi}$ and $\ave{\eta}$. The averages of the variables $\xi$
and $\eta$ have been defined as

\begin{eqnarray}
\ave{\xi} & = & \int e^{-2a_{1}t_{1}}P(t_{1}) \nonumber \\
          & = & \hat{P}_{1}(s=2a_{1}),
\end{eqnarray}

and

\begin{eqnarray}
\ave{\eta} & = & \int e^{-2a_{2}t_{2}}P(t_{2}), \nonumber \\
           & = & \hat{P}_{2}(s=2a_{2}),
\end{eqnarray}

where $\hat{P}_{i}(s)$ is the generating function (Laplace
transform) of $P_{i}(t)$. Laplace transforms of similar random
time intervals were considered by Godr\'{e}che and Luck
\cite{godreche} in the context of {\it renewal processes}. In such
stochastic processes {\it events} occur at the random epochs of
time $t_{1},t_{2}...$ from some time origin $t=0$. Change of sign
of the position of a random walker is one such example. The
intervals of time between those events are treated as independent
and identically distributed random variables, very similar to the
random duration of the primary processes in our case. Depending on
whether the mean of the probability distribution $P(t)$ is finite
or not, two different types of behavior are expected, as already
observed in  continuous fractal-time random walk \cite {shlesinger} and
in the case of renewal processes \cite{godreche}.

{\bf (i) Random application times with finite means:}

When the mean of the probability distribution is finite, we can
write the averages of the random variables $\xi$ and $\eta$ in
terms of a series

\begin{equation} \label{eq:}
  \ave{\xi} =  1 - 2a_{1}\ave{t_{1}} + \textrm{higher terms},
\end{equation}

and

\begin{equation} \label{eq:}
   \ave{\eta} = 1 - 2a_{2}\ave{t_{2}} + \textrm{higher terms}.
\end{equation}

In the scaling limit of $q \rightarrow 0$, {\it i.e.} small
$a_{i}$, the higher order terms can be neglected with the result

\begin{equation} \label{eq:}
\ave{\xi} =  1 - \nu_{1}q^{z_{1}}\ave{t_{1}} \quad \textrm{and}
\quad \ave{\eta} = 1 - \nu_{2}q^{z_{2}}\ave{t_{2}}.
\end{equation}

Again invoking the asymptotic limit $q \rightarrow 0$, the primary
process with smaller $z_{i}$ (dynamic exponent) dominates the
average SF and hence the scaling behavior of the roughness.
Therefore, as long as the probability distribution has a finite
mean, the scaling exponents corresponding to the average roughness
are identical to those of the uniform duration case, with the
average duration being the uniform period. In the next section,
this will be shown explicitly for various probability
distributions. We will also present results of numerical
simulations which shows the validity of our conclusion even if one
or both of the primary processes contain nonlinearity.

{\bf (ii) Random application times with diverging mean:}

In some specific cases of broad distributions the average of the
probability distribution diverges. One such example will be
distribution with a fat power law tail $t^{-(\mu + 1)}$ with $\mu
< 1$ (this is similar to the behavior found in continuous-time
random walks, giving rise to the so-called {\it fractal time}
behavior \cite{shlesinger}. In this case, $\ave{\xi}$ and
$\ave{\eta}$ in the scaling limit ($q \rightarrow 0$) are given by
\cite{godreche}

\begin{equation}
 \ave{\xi} \simeq 1 - \kappa _{1}(2a_{1})^{\mu} \quad \textrm{and} \quad
 \ave{\eta} \simeq 1 - \kappa _{2}(2a_{2})^{\mu}.
\end{equation}

By putting $a_{i} \sim q^{z_{i}}$, we obtain

\begin{equation} \label{eq:}
\ave{\xi} \simeq 1 - \kappa '_{1}(q)^{\mu z_{1}} \quad
\textrm{and} \quad \ave{\eta} \simeq 1 - \kappa '_{2}(q)^{\mu
z_{2}}.
\end{equation}

Hence, the dynamic exponents are modified as $\mu z_{i}$. The
process with smaller $z_{i}$ will still dominate and yield the
cyclical dynamic exponent $z_{new} = \mu \; min\{z_{1},z_{2}\}$.
Physically this modification is the consequence of the modified
relation between the time and the number of cycles. If we ask what
is the total time elapsed after $n$ applications of the two
primary processes the answer is as follows: As long as the mean
duration of one application is finite (i.e., $\mu>1$ for
distributions with an algebraic tail), the total time is
proportional to n times this mean duration. For power-law
distributions with $\mu <1$ and diverging mean, the dominating
contribution to the total time comes from a finite number of
applications, each of them lasting a time of the order of
$n^{1/\mu}$. This also implies that for such systems in which the
roughness grows with time as $t^{\beta}$ (where $\beta$ is the
growth exponent of the primary process with the smaller $z$), it
will grow as $n^{\beta/\mu}$ with the number of applications $n$.

After a large number of cycles, once the roughness becomes
saturated, the roughness exponent $\alpha$ of the grown
self-affine surface should not depend on the random application
times of the primary processes. As a result, the exponent $\alpha$
will remain unchanged from the corresponding uniform duration
case, which is nothing but the $\alpha$ of the primary process
with smaller $z$. This is consistent with our prior observation
that the growth exponent $\beta$ is modified as $\beta_{new} =
\beta / \mu$ , so that the scaling relation
$\frac{\alpha}{\beta_{new}} = z_{new}$ remains valid. So for $\mu
< 1$, the exponent $\beta$ changes continuously depending on the
values of $\mu$.

For two primary processes with durations having two different
values of $\mu$ of the power law distributions the new dynamic
exponent becomes $z_{new} = min \{ \mu_{1}z_{1},\mu_{2}z_{2} \}$,
and $\beta = \alpha / z_{new}$, where $\alpha = \frac{1}{2} \{
z_{new} - (d-1) \}$.
If one of the processes (say for $i=1$) has a finite mean while
that of the second ($i=2$) is diverging, the value of the
$z_{new}$ will be: $z_{new} = min \{z_{1},\mu_{2}z_{2}\}$, with
similar expressions for $\alpha$ and  $\beta$.
\subsection{Non-linear primary processes}

{\bf (i) Application times with finite mean:}

In this case we expect our approximate RG approach, introduced for
periodic processes and described in the previous section, to still
be applicable. The few initial RG iterations will coarse grain all
time-fluctuations smaller than a typical time of the order of the
average time period (the sum of the mean application times of the
two primary processes). Although the distributions will always
allow remaining fluctuations from cycles longer than the mean,
their probability decreases with their length and we do not
anticipate them to be relevant to the long time behavior.

{\bf (ii) Random application times with diverging mean:}

In this case the approximate RG cannot be applied since the
dominating effect is contributed from a few extremly long
applications, rather than from the accumulation of n applications.
This behavior strongly suggests that each of the primary processes
should be renormalized independently form the other and their
relative contributions should be compared based on their
fully-renormalized scaling behavior (contrary to the case (i)
above where the renormalization is carried on after the two
processes have been combined to a single effective process).

If that would be the case, the former result $z_{new} = min \{
\mu_{1}z_{1},\mu_{2}z_{2} \}$, will hold in general with the z(i)
of the two primary processes {\it whether they are linear or not}.
The exponent $\beta_{new}$ will be given by the larger of
$\beta(i)/\mu(i)$. $\alpha$ will simply be that of the dominating
process determining $z_{new}$.

{\bf (iii) Random application times with one finite and one
diverging mean:}

If one of the two processes (say the first one) has a finite mean
application time while it diverges for the other (the $2^{nd}$),
some special care is required. We can again imagine performing a
few initial RG iterations until the average duration of the first
process is of order one. The duration of the second process will
be rescaled by the same rescaling factor but will continue to be
dominated by the few extremly long applications.
If we have $z_{2} < z_{1}$ to begin with, then
$z_{new} = \mu_{2}z_{2}$ as before. If, on the other hand we begin
with $z_{1} < z_{2}$, then $z_{1}$ is unaffected during the
initial RG iterations and we have to compare the contribution from
n applications of the first process with the few extremly long
applications of the second one. Again, we expect the smaller of $z_{1}$ and
$\mu_{2}z_{2}$ will dominate and yields $z_{new}$ of the combined
process. In short these results amount to assigning the value
$\mu_{1}=1$ for the finite-mean process and using again the same
expressions for $z_{new}$ and $\beta_{new}$ in {\it (ii)} above.

\section{Numerical Simulations of Randomly Alternating Growth
with Different Application time Distributions}

Below we will consider different probability distributions $P(t)$
(both discrete and continuous), and try to determine the scaling
behavior using the SF (19). The variables defined in the previous
section will continue to be used. The average duration of the two
primary processes will be denoted by $T_{1}$ and $T_{2}$ for all
types of probability distributions (as long as the average exists
and finite).

We have carried out numerical simulation of various linear and
nonlinear discrete growth models to verify our results
corresponding to different probability distributions. The system
size used in our simulation was varied between 128 to 4096 lattice
units. Periodic boundary condition is employed to keep the finite
size effects to a minimum. A typical cycle consisted of average
deposition ($T_{1}$) of 8 layers (average number of particles per
site) and the same amount of average desorption ($T_{2}$). The
maximum number of cycles n was changed between 512-8192 to reach
the saturated roughness phase. The roughness data was taken for
$\sim 1000-6000$ independent runs to average over the realizations
of the random deposition noise as well as the probability
distributions for the random duration.

\subsection{Uniform Distribution}

The durations of the primary processes are equally probable
between two limits $T_{max}$ and $T_{min}$. Hence, the probability
density is $P(t) = \frac{dt}{T_{max} - T_{min}}$. This yields

\begin{equation} \label{eq:}
\ave{\xi} = \ave{\exp (-2a_{1}t_{1})}     \nonumber
\end{equation}

\begin{equation} \label{eq:}
= \frac{1}{T_{max} - T_{min}} \int_{T_{max}}^{T_{min}} \exp
(-2a_{1}t_{1}) dt_{1}
\end{equation}

\begin{equation} \label{eq:}
= \frac{1}{2a_{1}(T_{max} - T_{min})} [ \exp (-2a_{1}T_{min}) -
                                    \exp (-2a_{1}T_{max} ]
\end{equation}

\begin{equation} \label{eq:}
= \frac{\exp (-a_{1}T_{max}) \exp (-a_{1}T_{min})} {2a_{1}(T_{max}
- T_{min})} \bigg{[} \exp (a_{1}T_{max}) \exp (-a_{1}T_{min})
           -  \exp (-a_{1}T_{max}) \exp (a_{1}T_{min}) \bigg {]}
\end{equation}

\begin{equation} \label{eq:}
= \exp (-2a_{1}T_{1}) \frac{sinh \{2a_{1}(\Delta T_{1}/2)\}}
                               {\{2a_{1}(\Delta T_{1}/2)\}},
\end{equation}
where $T_{1} = (T_{min} + T_{max})/2$ is the mean time and $\Delta
T_{1} = T_{max} - T_{min}$. Similarly,

\begin{equation} \label{eq:}
\ave{\eta} = \exp (-2a_{2}T_{2}) \frac{sinh \{2a_{2}(\Delta
T_{2}/2)\}}
                               {\{2a_{2}(\Delta T_{2}/2)\}},
\end{equation}

Clearly, the uniform duration case is recovered when $\Delta
T_{i}$ vanishes.

In the scaling limit of small $q$,

\begin{equation} \label{eq:}
\ave{\xi} \sim \exp (-2a_{1}T_{1})  \quad \textrm{and} \quad
<\eta> \sim \exp (-2a_{2}T_{2}),
\end{equation}

since $\frac{sinh \{2a_{i}(\Delta T_{i}/2)\}}{\{2a_{i}(\Delta
T_{i}/2)\}} \sim 1$ as $a_{i} = q^{z_{i}} \rightarrow 0$. In this
limit, the SF reduces to

\begin{eqnarray}
S_{c}(q,n) \sim \frac{D_{c}}{a_{c}(q)} \left[1-\exp
\big(-2a_{c}(q) \, T \, n \big) \right],
\end{eqnarray}

where $T = T_{1} + T_{2}$ is the average duration for one complete
cycle with $T_{1} = pT$ and $T_{2} = (1-p)T$. The definitions of
$a_{c}$ and $D_{c}$ are same as the uniform duration case. Hence,
in terms of the mean durations of the primary processes, the above
SF is identical to the SF (10) of uniform duration. Thus the
scaling exponents corresponding to the mean cyclical roughness
will be identical to the scaling exponents of the primary process
with smaller $z$, as observed in the case of constant duration of
primary processes.

This may be seen in Fig. 1, where two linear processes namely EW
and DT (MH universality) growth models are used to simulate
cyclical growth with random durations distributed uniformly
between two values. The scaling exponents obtained ($\alpha = 0.50
\pm 0.02$ and $\beta = 0.25 \pm 0.03$) are EW exponents. However,
the mean roughness is slightly higher (Fig. 2) than the roughness
of the corresponding uniform duration case. This is expected as
$\frac{sinh \{2a_{i} (\Delta T_{i}/2)\}}{\{2a_{i}(\Delta
T_{i}/2)\}} > 1$ and hence from eqs. (31) and (32), $\ave{\xi} >
\exp (-2a_{1}T_{1})$ and $\ave{\eta} > \exp (-2a_{2}T_{2})$.

If we include nonlinear processes in our simulations the scaling
behavior still remain unchanged from the corresponding uniform
duration case. In Fig. 3, the mean roughness is plotted against
the number of cycles (for different system sizes) using two
nonlinear models, namely, the KK model (KPZ universality) and the
LD model (MBE universality). The durations of those primary
processes were taken to be uniformly distributed with identical
average. The result: $\alpha = 0.50 \pm 0.02$ and $\beta = 0.32
\pm 0.03$ is consistent with the KPZ values same as obtained in
the nonrandom case.

\subsection{Poisson Distribution}
The distribution function is given by $P(t_{i}) = \frac{\exp
(-T_{i}) T_{i}^{t_{i}}}{t_{i}!}$, with $T_{i}$ being the mean.
Averaging over this discrete distribution function

\begin{equation} \label{eq:}
\ave{\xi} = \sum_{t=0}^{\infty} \exp (-2at) \frac{\exp (-T_{1})
T_{1}^{t}}{t!}
\end{equation}

\begin{equation} \label{eq:}
= \exp \{ T_{1} (\exp (-2a_{1}) - 1) \},
\end{equation}

where $T_{1}$ is the mean duration for the first process.
Similarly, $\ave{\eta} = \exp (T_{2} (\exp(-2a_{2}) - 1)$ with
$T_{2}$ being the average duration of the second process.

As before, to extract the scaling exponents small $q$, i.e. small
$a_{i}$ ($\sim q^{z_{i}}$), limit is taken. In this limit
$\ave{\xi} \sim \exp \{T_{1} (1 - 2a_{1} - 1)\}$ =
$\exp(-2a_{1}T_{1})$, as obtained in the uniform duration case.
Hence the scaling exponents corresponding to the average roughness
will be same as the uniform duration case.

\subsection{Exponential Distribution}
The distribution function $P(t_{i}) = \frac{1}{T_{i}}
\exp(-t_{i}/T_{i})$, where $T_{i}$ is the mean value. This yields

\begin{equation} \label{eq:}
\ave{\xi} = \frac{1}{1+2a_{1}T_{1}} \quad \textrm{and} \ave{\eta}
= \frac{1}{1+2a_{1}T_{2}},
\end{equation}

with $T_{1}$ and $T_{2}$ being the average duration of the two
primary processes. In the asymptotic scaling limit $q \rightarrow
0$, $\ave{\xi} \sim 1-2a_{1}T_{1}$ and $\ave{\eta} \sim
1-2a_{1}T_{2}$. Since $a_{i} \sim q^{z_{i}}$, the process with
smaller $z_{i}$ will dominate the SF (19). Hence, the scaling
exponents are once again identical to those of the non-random
case.

\subsection{Power-law Distribution}

If the long time behavior of a primary process has a fat tail,
then duration of that process can be taken to be distributed as a
power law of time with

\begin{eqnarray}
          P(t) & = & \mu t^{-(\mu + 1)}  \quad   t \ge 1
\nonumber \\
              & = & 0,          \quad            t < 1
\end{eqnarray}

where $\mu$ is the only parameter of the distribution. Average
duration is easily calculated for the above {\it p.d.f.}

\begin{eqnarray}
  \langle t \rangle = \int _{1}^{\infty} P(t) dt
      = \mu \int _{1}^{\infty} t^{-(\mu + 1)} dt
      = \frac{\mu}{1-\mu} t^{(1-\mu)} |_{1}^{\infty}.
\end{eqnarray}

There are two possibilities:

(i) For $\mu > 1$: $\ave{t} = \frac{\mu}{\mu-1}$. Hence the
average and all the higher moments are finite.

(ii) For $\mu \le 1$: $ \ave{t} \rightarrow \infty$, the average
diverges.

Below we will consider two possibilities separately.

\subsubsection{$\mu > 1$}

In the scaling limit $q \rightarrow 0$, the averages of the random
durations $\xi$ and $\eta$ are given by (see eq. (23))

\begin{equation}
 \ave{\xi} \simeq 1 - 2a_{1}T_{1} \quad \textrm{and} \quad
 \ave{\eta} \simeq 1 - 2a_{2}T_{2}.
\end{equation}

From the expression of the SF (19) we can see that the process
with smaller $z$ will dominate the average SF and hence the
roughness. Hence, the scaling exponents should not change from
their corresponding uniform duration case.

We simulated MH deposition and EW desorption alternately, with
their duration distributed with a power law tail ($\mu > 1$). Fig.
4 shows the scaling of the average roughness with number of cycles
for different values of $\mu > 1$. The exponent $\beta \sim 0.25$
for all those cases, which is the $\beta$ of the dominating EW
process. The roughness exponent $\alpha$ should not change from
its uniform duration value and we obtain the EW exponent $\alpha$
($= 0.48 \pm 0.05$) in our simulations.

\subsubsection{$\mu < 1$}

In section III we have already analyzed this case of diverging
mean duration. The growth exponent $\beta$ is supposed to change
continuously depending on the value of the parameter $\mu$. This
behavior was observed (Fig. 5) in our simulation of the cyclical
process combining (1+1)d EW and MH model. Random durations of both
the primary processes obey power law distributions with identical
$\mu$. The roughness exponent $\alpha$ retains its value (Fig. 6)
corresponding to uniform application time ({\it i.e.} the EW
value).

Presence of nonlinearity in one or both the primary processes did
not change the essential scaling behavior from the linear case. We
simulated nonlinear KK (KPZ) and LD (MBE) models having durations
distributed with power law tails (both $\mu > 1$ and $\mu <1$).
For $\mu > 1$, no difference was observed in the scaling exponents
from the corresponding uniform duration case. For $\mu <1$,
however, the growth exponent $\beta$ increased continuously as the
parameter $\mu$ decreased to zero, consistent with the scaling
behavior found for the linear primary processes. The roughness
exponent $\alpha$ remain unchanged in all those cases.

Therefore, for intermittent applications with a diverging average,
the dynamic exponent and the growth exponent start to differ from
those of the uniform case whereas the roughness exponent remains
the same.

\section{Conclusions}

  We are now in a position to provide the answers to the two central
questions we posed at the beginning of this paper:

1. To what extent the cyclical growth behavior describes also
randomly alternating processes? - The scaling behavior of the
perfectly periodic growth processes continues to hold for random
application times as long as the latter have finite means. Small
fluctuations in the application times are thus irrelevant. For
linear systems it means that the primary process with the smallest
dynamic exponent will dominate. For non-linear processes, the
approximate RG procedure devised for cyclical processes may still
be applied to extract the asymptotic scaling behavior.

2. When cyclical growth fails to describe randomly alternating
processes, what are the their new scaling properties?
 If at least the distribution of one of the application time decays as a power
law $t^{-1-\mu}$ with $\mu<1$ (such that its mean diverges), the
effective dynamic exponent is reduced by a multiplicative factor
$\mu$. For linear processes it definitely implies that the process
with the smallest effective dynamic exponent dominates (the value
of $\mu=1$ should be assigned for processes with a finite mean).
We claim that the same continues to hold for non-linear processes
as well. This claim follows from the fact that a primary process
with a diverging mean application time is dominated by a few
especially long applications
and therefore renormalizes independently from the other primary
process. Theoretical studies to further confirm this behavior will
be most welcome.

As far as experimental investigations are concerned, they should
be straightforward for systems with finite means (e.g., by methods
similar to those used to measure the scaling behavior of cyclical
growth). It will be more challenging to conduct experimental
investigations of surface growth by randomly alternating processes
where at least one of them has a diverging mean application time.

\section{Acknowledgments}

This research was supported by the ONR under grant No.
N00014-00-1-0057.

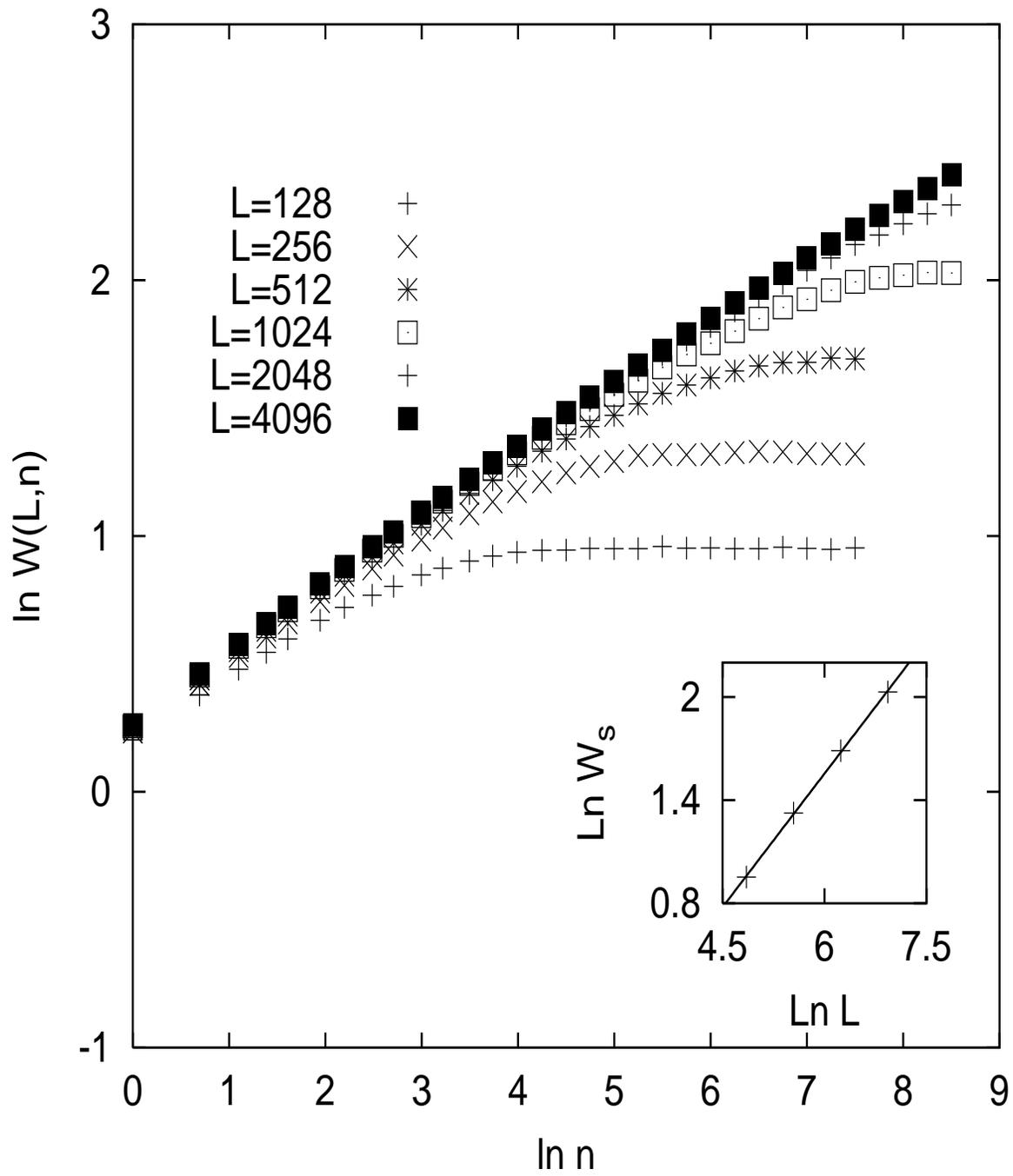
\begin{figure}
\caption{$\ln W$ (roughness) {\it vs} $\ln n$ (number of
applications) of the MH/EW process with finite-mean random
application times [Inset: $\ln W_{s}$ (maximal roughness) {\it vs} $\ln
L$].} \label{Fig. 1}
\end{figure}

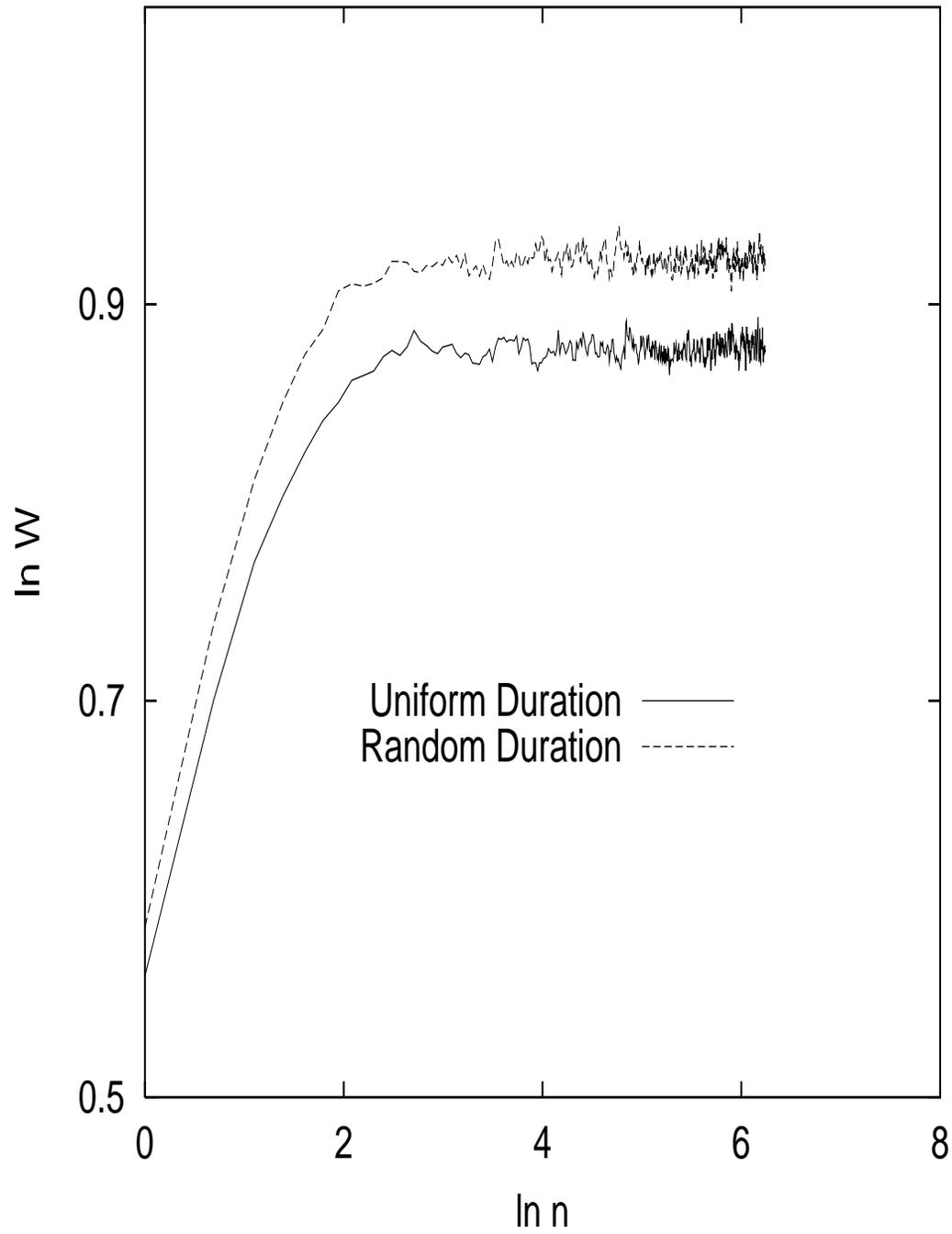
\begin{figure}
\caption{The roughness $W$ {\it vs} number of applications  $n$ of
the MH/EW process with finite-mean randomly
alternating durations, compared with the uniform
({\it i.e.} cyclical) case ({\it log-log} plot).} \label{Fig.
2}
\end{figure}

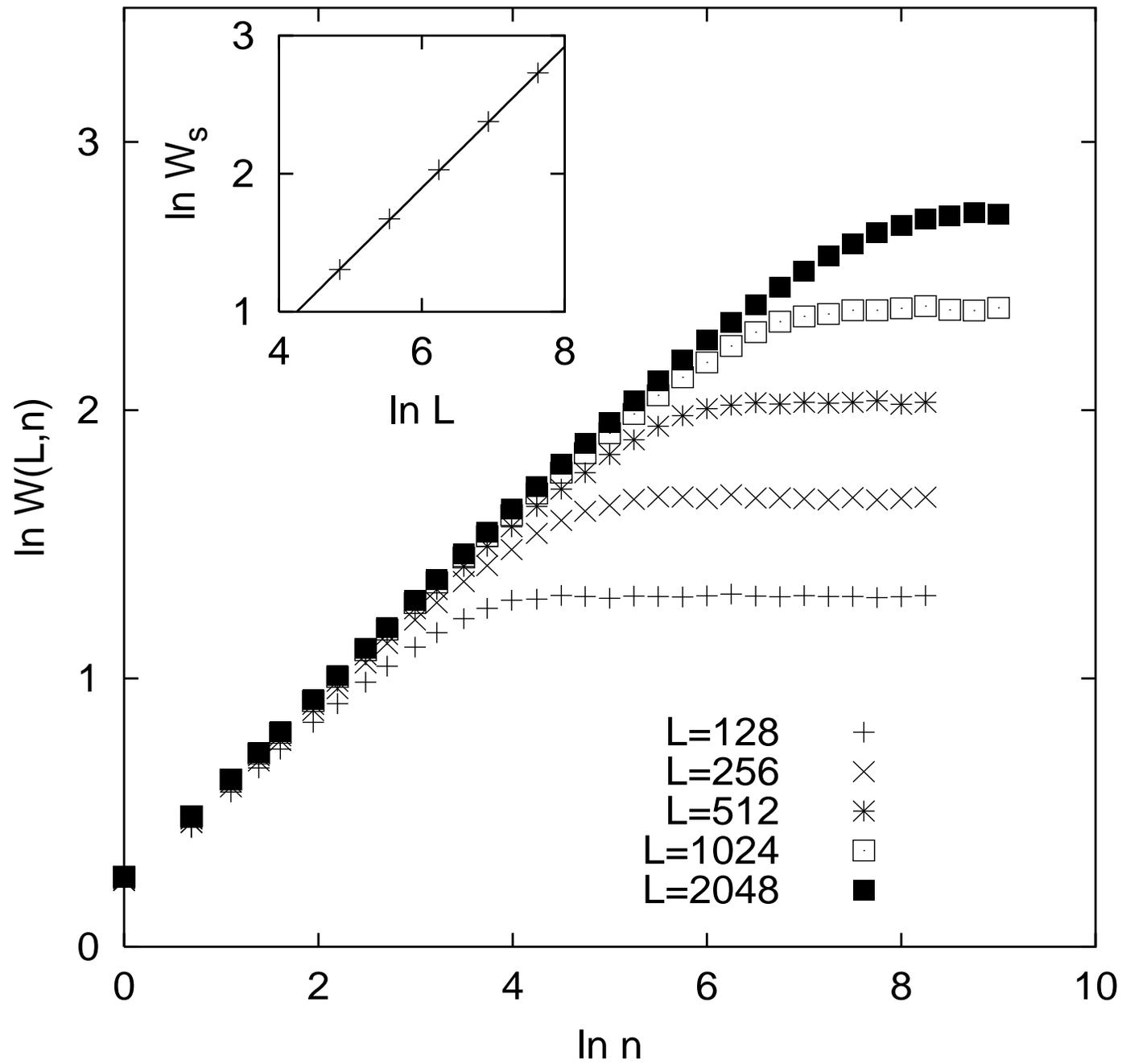
\begin{figure}
\caption{$\ln W$ (roughness) {\it vs} $\ln n$ (number of
applications) of the randomly alternating MBE/KPZ process for
different system sizes $L$ [Inset: $\ln W_{s}$ (maximal
roughness) {\it vs} $\ln L$].} \label{Fig. 3}
\end{figure}

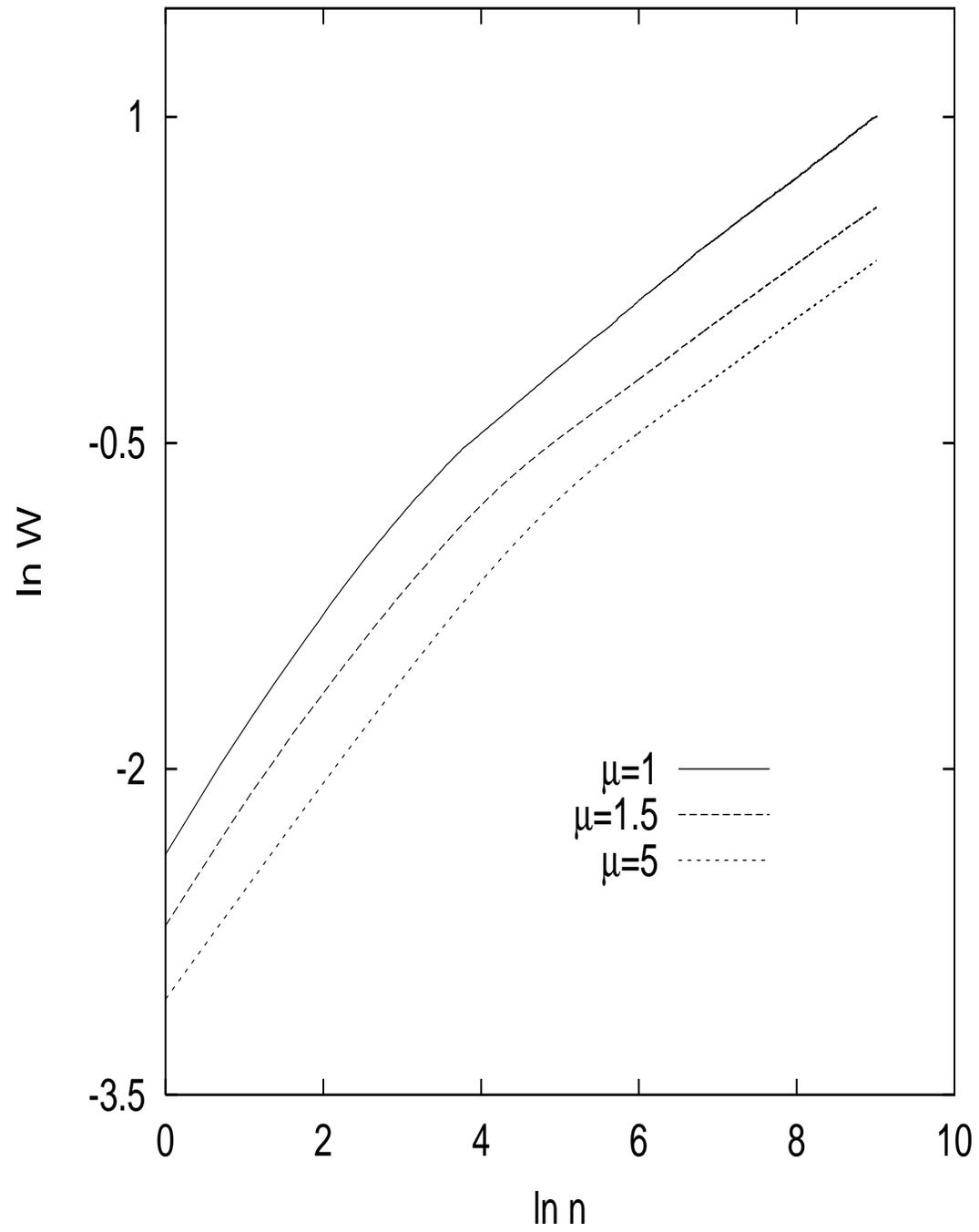
\begin{figure}
\caption{$\ln W$ {\it vs} $\ln n$ of the power-law distributed
MH/EW randomly alternating process for different values of $\mu > 1$.}
\label{Fig. 4}
\end{figure}

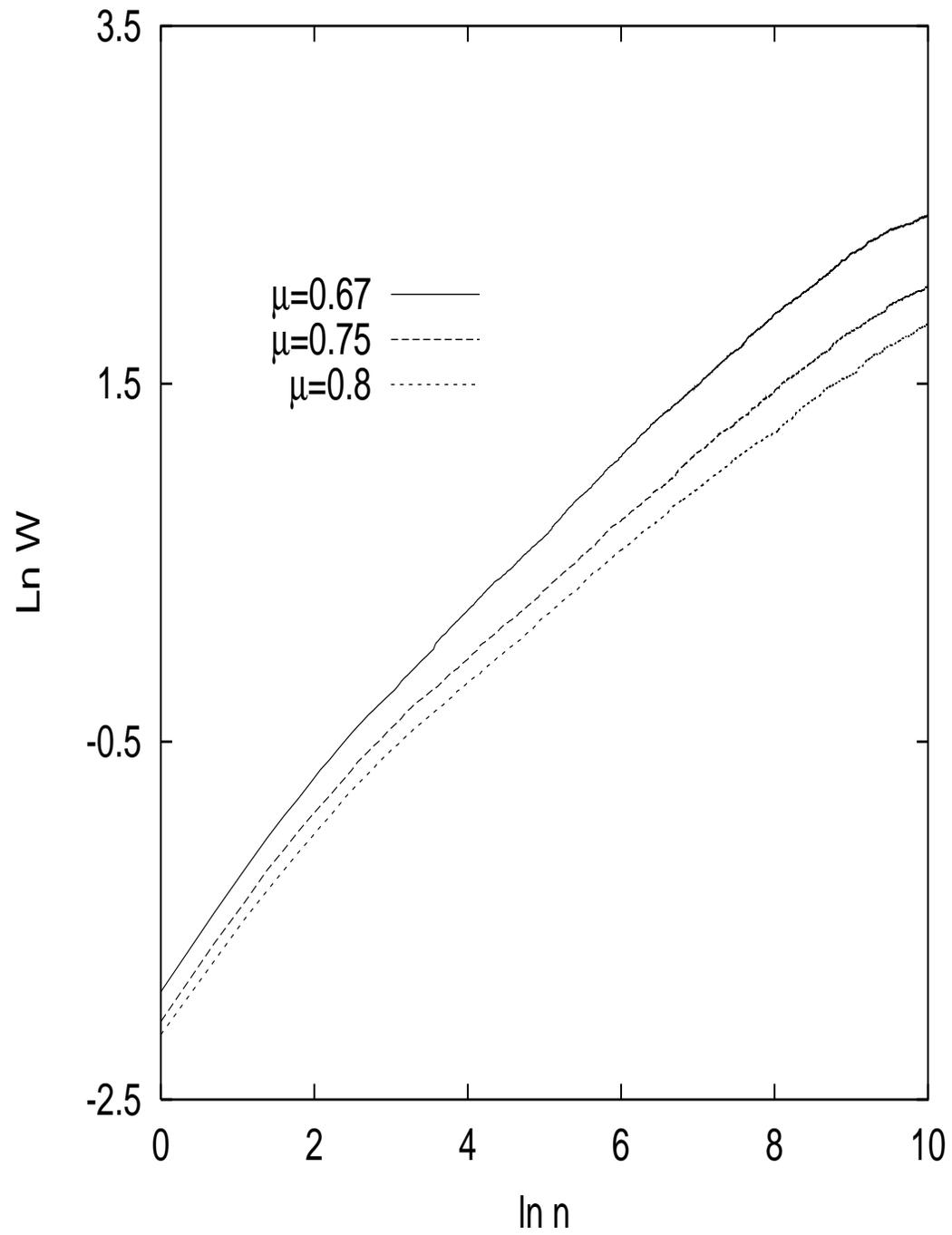
\begin{figure}
\caption{$\ln W$ {\it vs} $\ln n$ of the power-law distributed
MH/EW randomly alternating process for different values of $\mu < 1$.}
\label{Fig. 5}
\end{figure}

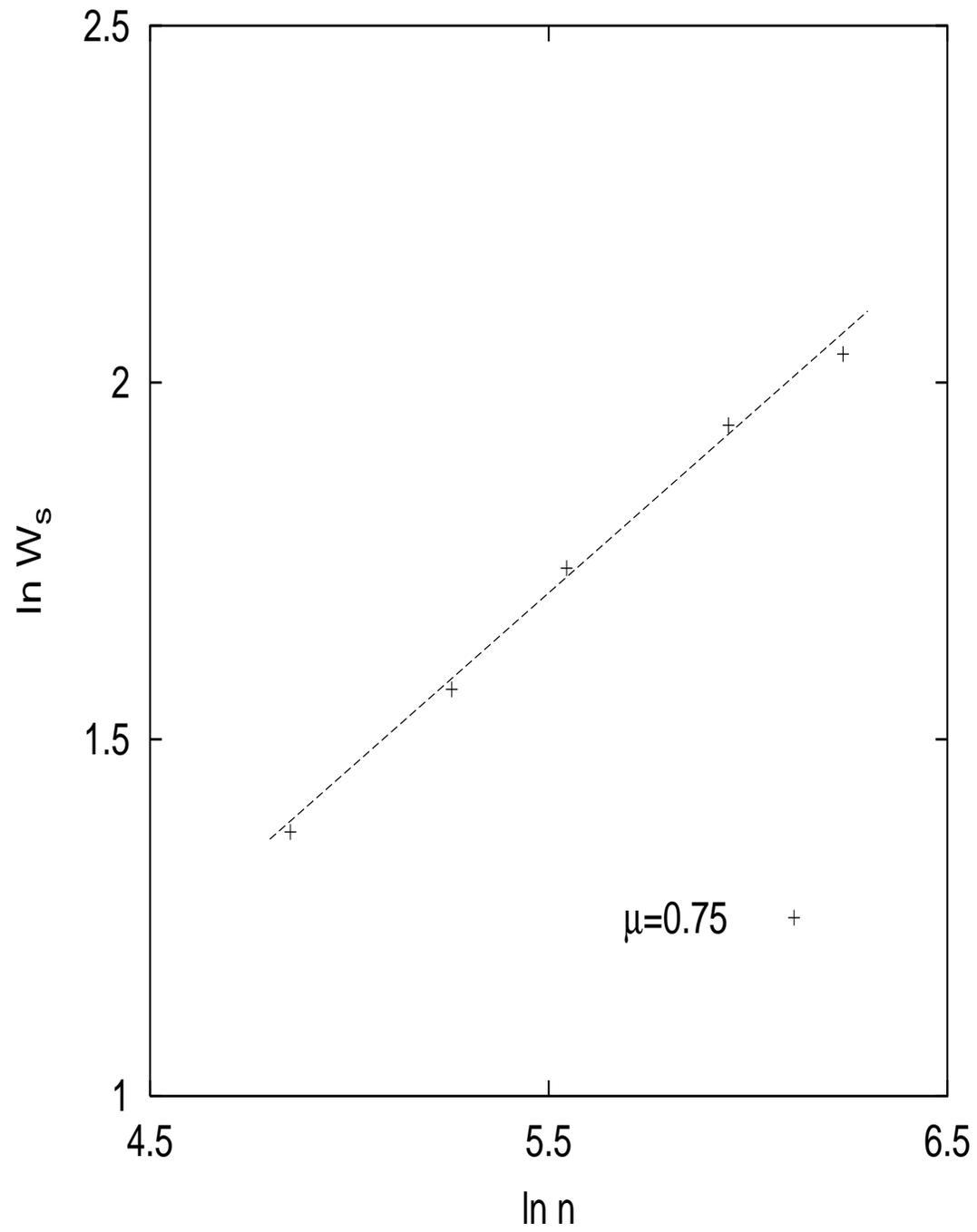
\begin{figure}
\caption{The roughness $W$ {\it vs} number of applications $n$ of
the power-law ($\mu=0.75$) distributed MH/EW process for
different system sizes $L$ ({\it log-log} plot) [Inset: $\ln
W_{s}$ (maximal roughness) vs $\ln L$].} \label{Fig. 6}
\end{figure}

\end{document}